\documentclass[conference]{IEEEtran}
\IEEEoverridecommandlockouts
\usepackage{cite}
\usepackage{amsmath,amssymb,amsfonts}
\usepackage{algorithmic}
\usepackage{graphicx}
\usepackage{textcomp}
\usepackage{xcolor}
\usepackage{subfigure}
\usepackage{color}

\def\BibTeX{{\rm B\kern-.05em{\sc i\kern-.025em b}\kern-.08em
    T\kern-.1667em\lower.7ex\hbox{E}\kern-.125emX}}
\begin{document}

\title{Measuring Security in 5G and Future Networks}

\author{\IEEEauthorblockN{Loay Abdelrazek\IEEEauthorrefmark{1}, Rim ElMalki\IEEEauthorrefmark{2}, Filippo Rebecchi\IEEEauthorrefmark{2}, Daniel Cho\IEEEauthorrefmark{1}}
\IEEEauthorblockA{\IEEEauthorrefmark{1} Standards \& Technology, Ericsson, Sweden. Email: \{name.surname\}@ericsson.com}
\IEEEauthorblockA{\IEEEauthorrefmark{2} Standards \& Technology, Ericsson, France. Email: \{name.surname\}@ericsson.com}

}

\maketitle

\begin{abstract}
In today's increasingly interconnected and fast-paced digital ecosystem, mobile networks, such as 5G and future generations such as 6G, play a pivotal role and must be considered as critical infrastructures. Ensuring their security is paramount to safeguard both individual users and the industries that depend on these networks. 
An essential condition for maintaining and improving the security posture of a system is the ability to effectively measure and monitor its security state. In this work we address the need for an objective measurement of the security state of 5G and future networks. We introduce a state machine model designed to capture the security life cycle of network functions and the transitions between different states within the life cycle. Such a model can be computed locally at each node, or hierarchically, by aggregating measurements into security domains or the whole network. We identify three essential security metrics -- attack surface exposure, impact of system vulnerabilities, and effectiveness of applied security controls -- that collectively form the basis for calculating the overall security score. %With this approach, it is possible to provide a holistic understanding of the security posture, laying the foundation for effective security management in the expected dynamic threat landscape of 6G networks. 
Through practical examples, we illustrate the real-world application of our proposed methodology, offering valuable insights for developing risk management and informed decision-making strategies in 5G and 6G security operations and laying the foundation for effective security management in the expected dynamic threat landscape of 6G networks.
\end{abstract}


\section{Introduction} \label{sec:intro}

The ability to measure and quantify the cyber security posture is expected to be one key demand from 6G operators, as it allows them to make informed decisions in risk management, achieve compliance with regulatory requirements, and establish trust in the security of 5G and 6G networks~\cite{coronado2022zero}. 

With the first 6G deployments expected to happen sooner than 2030 \cite{ericsson6gspectrum}, the need to understand and mitigate the cyber risk environment will present unique challenges due to some unavoidable trends. Among those, the tendency of a more distributed 6G architecture, that is due to the scale and heterogeneity of data is  expected to go up by a factor of $10$ when compared to 5G. Finally, the ongoing trend towards open networks will materialize into a system consisting of multiple stakeholder domains, where mutual trust is not always accounted for or possible as the multiple parties can be involved in the service delivery (e.g., a service can be provided by a service provider leasing virtual resources from a cloud provider having its infrastructure provided by an infrastructure provider).

These factors, along with the technology evolution and capabilities embedded in systems, underscore the essential role of measurement in driving automated security management and providing a quantifiable assessment of the overall security posture of a deployment. Essentially, complexity and challenges are driving how security management and orchestration is performed, shifting from the current model based on static assessments and focused on compliance that is mostly executed offline, to a runtime security model that is constantly reevaluated due to the system dynamics towards a DevSecOps mode of operations. Within this landscape, automation becomes no longer optional. 

The integration of advanced security automation, reducing human intervention with zero-touch configuration, and simplifying security management via intent-based security are essential helpers in addressing the complexity of 6G networks, as it empowers network operators to actively manage the perceived threat level. Furthermore, the adoption of AI and ML techniques enhances automated operations at scale, enabling self-configuration, self-optimization, self-organization, and self-healing capabilities. These capabilities, underpinned by security measurements, not only streamline the workload of human operators but also minimize the risks of mis-configuring the network, thereby bolstering the resilience and security.

In this paper we investigate the key problem of developing a quantifiable \textit{score} that can serve to  assess effectively the overall security posture of 5G and 6G deployments. This score is expected not only to drive security automation and intent-based management, but also to serve as a comprehensible indicator of the current security posture, facilitating communication between technical and non-technical stakeholders. 

In our approach, we develop a more comprehensive model that considers the potential entry points to a 5G and 6G mobile network systems (i.e attack surface), the individual vulnerabilities of components (these can be either software, protocol, and configuration related), and the effectiveness of the applied security controls. The objective is to provide an objective holistic understanding of the security posture, to enable proactive risk management and informed decision-making in support of mobile network operations.

In this paper we have the following contributions:
\begin{itemize}
    \item Introduce a comprehensive finite state machine model to capture the security life cycle of network functions in mobile networks and provide a structured method to understand the security state transitions.
    \item Use the defined finite state model to propose and define three different security metrics that yield intrinsic for the objective calculation of a security state score.
    \item Propose a mathematical approach to calculate the network security posture based on the proposed metrics.
    \item Provide a holistic evaluation of a security posture in a mobile network by measuring three security metrics, Attack Surface Level, Vulnerability Impact Level and Security Control effectiveness.
\end{itemize}

The rest of the paper is organised as follows. Section~\ref{sec:state} introduces a security state machine model that is used to provide a systematic approach to identify and define the security metrics. Section~\ref{sec:metrics} discusses the main security metrics that are identified to constitute the overall score, illustrating also how they are calculated. Section~\ref{sec:related} addresses related works and positions our work. Section~\ref{sec:usage} provides examples of how to utilize the security metrics in 6G. Section~\ref{sec:conclusion} highlights some potential research directions, and draws conclusions.

\section{Security State Machine Model} \label{sec:state}

To identify the relevant security metrics and to expand on the existing work, we leveraged a Finite-State Machine (FSM) model to create a security state machine model. The security state machine model is used to identify the different states that affect the security posture of a network function. FSM provides an understanding of the changes in the security posture as a result of changes in the environment of the network function or the surrounding environment. Moreover, we define hierarchical levels of security states which represents the different layers where measurements can be collected and calculated, this hierarchy provides granularity and explainability to understand which parts of the networks and which network functions having a poor security state that needs enhancement, and their impact and contribution to the overall network security posture.

\subsection{Hierarchical Security Levels}

\begin{figure}[h!]
    \centering\includegraphics[width=0.35\textwidth]{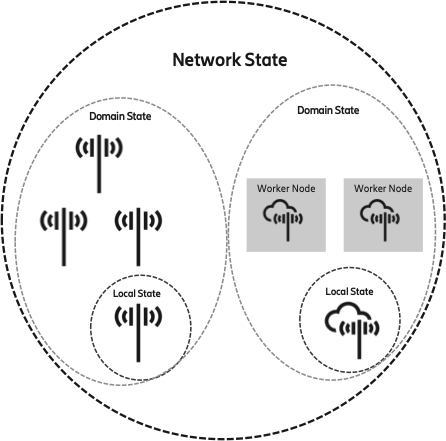}
    \caption{Security State Hierarchy Levels}
    \label{fig:state_levels}
\end{figure}

Security states can be measured and represented in a hierarchical level. As illustrated in Figure \ref{fig:state_levels}. a network can have three levels of security states where one is the subset of the other. In the foundational level is the Local Security State, which presents the security state of a single network node, for example a network node can be a Next Generation NodeB (gNB), a Central Unit (CU) or a Radio Distributed Unit (DU), etc. This only presents a singular enclosed view on the posture of the network node, no environmental factors impact its security state. 

The second level is the Domain Security State, it represents the security state of a group of network functions that are grouped into one a domain that implements a security policy and is administered by a single authority. The domain state will be the group of the local security states in one domain. For example, in a physical deployment, it can present the group of gNBs that are in the same geographical area, or are defined to be in the same security domain where a set of security policies and requirements are applied to this group of gNBs. In contrast, in a cloud deployment, a domain can be the worker node(s) that are forming a cluster of the applications. The last level is the Network Security State, which represents the state of the whole network, which is a aggregation of the states of the existing domains.

\subsection{Security Finite State Model}
\begin{figure}[t!]
    \includegraphics[width=0.8\linewidth]{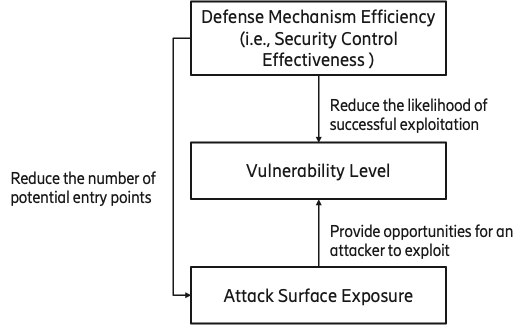}
    \caption{Metrics Relationships}
    \label{fig:metric_relation}
\end{figure}

Identifying the different security states a network node go through, provides an understanding on the possible events that transits from one state to another. This in turn assists defining the security metrics that are relevant to the observed states and allow to calculate a security posture score. Additionally, the security state machine model helps in understanding the relationships and dependencies between the metrics, if any. An example of the relationships of the identified metrics in this paper is illustrated in Figure \ref{fig:metric_relation}.

\begin{figure}[h!]
    \includegraphics[width=0.5\textwidth]{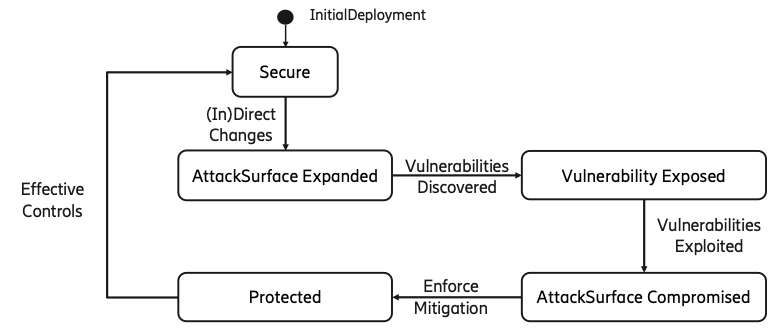}
    \caption{Security State Machine Model}
    \label{fig:security_state}
\end{figure}

As shown in Figure \ref{fig:security_state}, the states are identified as the nodes in the model, the states can be summarized as \textit{Secure, Attack Surface Expanded, Vulnerability Exposed, Attack Surface Compromised, and Protected}. In the initial deployment of a mobile network function (e.g. RAN Central Unit) it is assumed to be in a secure state since it is following the security design rules, hardened according to security benchmarks and baselines, scanned for vulnerabilities and threats have been identified and the mitigation control that matches the initial threat assumptions are configured. The issue lies during the operation phase after the network function is deployed and changes occur either to the network or to the underlying infrastructure (e.g., operating system, containers, container orchestrator, etc..). These changes lead to transitioning from an initial secure state to an attack surface expansion state. The changes to the network function can be related to new configurations, or any other change that is in the scope of the network function whether at the application level (e.g, adding a new feature) or at the system level (e.g., adding new libraries, or new capabilities such as deploying the network function on a container), etc. Moreover, the changes could be indirect and are induced by the surrounding network, for example a new User Equipment (UE) attached and connected to a cell is considered a change as it affects the attack surface, a new cell added to gNB or a DU added to the CU or any other topology changes to the network. Moreover, new attacks that have been detected in the network might have an impact on the security state of a network function as it will require the network function to evaluate itself and measure its posture to gain an understanding if it is impact by the vulnerability used in the attack or if its attack surface is exposed. The previous examples illustrate possible changes that could occur and which contribute to expanding the attack surface of a network function. Evaluating the entry points are essential, first to understand the causal effect of the changes and their contribution to increase or decrease the entry points. Second, to evaluate the possible entry points an attacker can leverage to exploit and compromise the network function.

Once the entry points of a network function have been identified after a change, they are evaluated to depict if there are any vulnerabilities that can be misused through the attack surface that has been identified. Hence, transitioning from just expansion in the attack surface to vulnerability exposure state. An important aspect is to evaluate if the vulnerability can be exploited, and what is the impact of the vulnerability in case of exploitation. If the vulnerability is exploitable then this is more likely to lead to a compromise of the system through the attack surface, which in return transfers the state from exposure of vulnerabilities to a compromised state.

Lastly, identifying and enforcing the suitable mitigation controls will achieve the transition between a compromised state of a network function to a protected state, where if the mitigation controls are correctly and effectively configured will lead that the network function return to a secure state that aligns with the security requirements set for the network function or the network as a whole.

\section{Security Metrics} \label{sec:metrics}

The National Institute of Standards (NIST) defines in \cite{schroeder2022performance} security metrics as quantifiable measurements that are used to provide an understanding on the security status and posture of a system of service through the collection and analysis of the relevant data to what is being measured. In this paper, we define the metrics, and provide the mathematical models to calculate them. 

The identified security metrics in this paper can be aggregated to eventually calculate the score of a security state, or they can use individually for separate goals and requirements. For example, it could be of interest for one mobile network to measure the vulnerability impact level in the network functions in order to prioritize which network function and which vulnerability in the network function can be patched first. Another example, could be that it is of interest to set separate goals to the required threat level in part of the network or on a specific network function. This modularity of the definition and usage of metrics allows %several opportunities specially 
to set separate requirements in the network resulting in more understanding of the security shortcomings. In addition, modularity  assists on facilitating a risk-based decision where cost is taken into consideration to invest in a security mitigation solution based on the prioritized risk.

In this section, we describe in details the security metrics defined as a result of the FSM described in section \ref{sec:state}. Additionally, we provide an example illustrating how to calculate the security control effectiveness of radio security controls that protect the radio control plane stack. Moreover, we provide another practical example of how to calculate a mis-configuration vulnerability score for a Radio Central Unit Control Plane Network Function (CU-CP) to achieve the purpose of automatically patching vulnerabilities that are induced by human errors.

\subsection{Security Control Effectiveness Metric}

Security control effectiveness refers to the degree to which a security control successfully achieves its intended purpose of protecting, mitigating or reducing security risks or threats within a system. It essentially measures how well a security control performs in protecting assets, systems, or data from potential security incidents or breaches.

The security control effectiveness metric depends on two measures: \textit{security control coverage} and \textit{security control correctness}. Security control coverage determines if the necessary controls exist (e.g., a firewall has been deployed), and security control correctness measures how well the controls have been implemented (e.g., the firewall rules are effective).

Security control effectiveness metric, which is denoted as $ScE$, is the dot product of the security control coverage (${SC_{CV}}$) and security control correctness (${SC_{CR}}$) measures:

\begin{equation}
ScE={SC_{CV}}\cdot{SC_{CR}}
\end{equation}
 
${SC_{CV}}$ represents the availability of controls that satisfy a security requirement. It follows binary associations to the security controls that are implemented. The result is a vector of binary values depending on the availability of the control. For example, if there is a security requirement concerning the radio interface to ensure a high level of protection on the radio control plane stack, it translates into ensuring confidentiality and integrity protection, as recommended by \cite{3gpp33501}. Thus, to fulfill the requirement both encryption and integrity algorithms need to be configured. As an example, if one of the controls is only implemented then:

\begin{equation}
\overrightarrow{SC_{CV}}=[SC_{Encryption}, SC_{Integrity}]
\end{equation}

The previous equation  will then lead to a score of $0.5$ since  only one control was implemented, e.g., if the encryption control was implemented then the vector will be: 

\begin{equation} \overrightarrow{SC_{CV}}=[0, 1]\end{equation}

In this security requirement example, the security control effectiveness will be equal to ${ScE}$ $=$ ${SC_{CV}}$ since cryptographic algorithms are implemented as a preference list, where the preference does not usually reflect the strength of the cryptographic algorithm, but rather the impact of the used algorithms on the underlying infrastructure specially if encryption is handled in software \cite{khan2018aes}\cite{ekdahl2021snow}. Thus, ($SC_{CR}$) will be equal to $1$. However, if it has been found that the Null encryption scheme \cite{3gpp33501} exists in the algorithms list and is considered as the first choice, then this will impact the correctness of the controls and a penalty score ($P_N$) will be deducted from $SC_{CR}$. %The preference list is used to negotiate with the UE the encryption capabilities that shall be applied based on the capabilities provided by the UE to the network and the preference list that is configured on the radio network function. Thus, if the UE supports a Null scheme, and the radio network function has the Null scheme as first on its preference list, then it will be the choice that will be agreed on. 
The risk is increased in this situation as no encryption or integrity protection will be applied on the communication link. The penalty score that will be deducted is based on the context of the security control, in the context of radio related controls, the penalty score is a factor of the number of UEs that are currently connected in a cell. This is attributed to the fact that the security configuration of the cell is a global configuration and it will impact all UEs that will attach and connect to the cell. A penalty score can be realized as shown below:

\begin{equation}
P_{N}=\frac{UE_{Conn}}{UE_T}, \quad P_{N}\in [0,1]
\end{equation}

where $UE_{Conn}$ represents the number of connected UEs in a cell at time $t$, and $UE_T$ is the total number of UEs that can connect to a cell, in other words the capacity of a cell in terms of connected UEs. Thus, if there are $100$ UEs connected to a cell that is designed with max capacity of $300$ connected UEs, then the penalty value can be calculated as follows:

\begin{equation}
Penalty = {SC_{CR}}\times{P_N}
\end{equation}

This equation will result in a penalty value of $0.33$, and the final security control correctness ($SC_{CR}$) becomes $0.66$ which is derived from the below:

\begin{equation}
{SC_{CR}} = {SC_{CR}} - Penalty
\end{equation}

Finally, the ($ScE$) score will be the dot product of $0.5$ and $0.66$ which will result in a score of $0.33$. This can be interpreted as that the effectiveness of the radio security controls is only $30\%$, thus the remaining $70\%$ represents the shortcoming or the gap of the current mitigation control to fulfill the security requirement which was the full protection of radio control plane stack.

This metric can be used to evaluate the network slice security state, specifically to check that effective isolation between slice traffic is consistent with predefined policies and requirements. 

\subsection{Vulnerability Level Metric}

A security vulnerability refers to a weakness or flaw in a system, network, software application, or hardware device that could be exploited by an attacker to compromise the security of the system and conduct a successful attack. Vulnerabilities can be caused by design flaws, implementation errors, mis-configurations, or other factors. The presence of vulnerabilities increases the probability of expanding the attack surface, thereby providing attackers with opportunities to compromise a system. In this paper we further categorize Vulnerabilities as the below:

\begin{itemize}
\item \textbf{Software vulnerabilities:} describe vulnerabilities in the software stack, including the applications that run the telecommunication functionalities, cloud-native services, operating system and container images and possible vulnerabilities towards AI/ML models.
\item \textbf{Protocol vulnerabilities:} describe vulnerabilities in the network protocol stacks on the different interfaces, for example 3GPP interfaces (IP , Radio) and OAM interfaces.
\item \textbf{Configuration vulnerabilities:} describe vulnerabilities that are introduced due to human error that results in mis-configurations.

\end{itemize}

A vulnerability metric measures the impact and exploitability of vulnerabilities of a specific entry point of the system and it facilitates priority-based remediation decisions (e.g., patching of software). Where each of the previous category can presents its own standalone metric. Thus, a total score of vulnerability metric can be divided into three sub-metrics as defined above.

GSMA \cite{gsmathreats} has identified human threat as one of the main imminent threat to a mobile network, this has also been highlighted in previous reports by GSMA which highlights the impact of human induced errors that are manifested in mis-configurations. In this paper we try to present how to leverage the defined vulnerability metrics in order to provide a solution to prioritize and automate the patching of mis-configurations in future 6G mobile networks. Automated configuration vulnerability management is a use case that can benefit from the vulnerability metric, specifically mis-configuration vulnerabilities metric (i.e., based on human errors). It patches mis-configuration vulnerabilities in the system, in a timely and effective manner, taking into consideration the context of the patched network function based on its criticality, impact of vulnerability, in order to facilitate priority-based patching. 

To calculate the mis-configuration vulnerability metric, a number of relevant measures can be identified and defined, mainly, the ratio  of non-compliant rules ($R_{NC}$), vulnerability impact ($V_{Imp}$),  asset criticality ($A_{Cr}$), duration of compliance ($D_{NC}$), and environmental impact ($Env_{Imp}$). This metric can be calculated for a single network function (i.e., local security state), for a group of network functions that are grouped into one domain which implements a security policy and is administered by a single authority (i.e., domain state), or for the whole network which is a combination of the states of the existing domains (i.e., network state):

\begin{itemize}
\item For the local security state, the weights of the asset criticality and environment impact measures are equal to 0 since the contribution/impact of the surrounding environment is not considered/relevant. To obtain the mis-configuration vulnerability metric for the local security state ($VulMet_{L}^{Human}$), the three remaining measures can be averaged: 
\begin{equation}
VulMet_{L}^{Human}=\frac{R_{NC}+V_{Imp}+D_{NC}}{3}
\end{equation}
\item For the domain state, mis-configuration vulnerability metric ($VulMet_{D}^{Human}$) can be calculated by first averaging the five identified measures for each network function ($x$) and then averaging the resulting for all network functions ($X$) in a domain:
\begin{equation}
VulMet_{D}^{Human}=\frac{\sum_{1}^{X} \frac{R^x_{NC}+V^x_{Imp}+D^x_{NC}+A^x_{Cr}+Env^x_{Imp}}{5}}{X}
\end{equation}
\end{itemize}

The derivation of each measure is detailed in the following.

\textbf{Ratio of non-compliant rules measure:} 
This measure ($R_{NC}$) represents the ratio of non-compliant rules ($N_{NC}$) out of the total number of rules ($N_{TR}$) corresponding to a single network function. One possible example of calculating this measure is through an equation that can be defined as follows:
\begin{equation}
R_{NC}=\frac{N_{NC}}{N_{TR}},
\end{equation}

The value of $R_{NC}$ ranges between $0$ and $1$, $0$ being the desired value and $1$ being the worst value.

\textbf{Vulnerability Impact:} 
This measure ($V_{Imp}$) represents the vulnerability for a single network function. It can be obtained using the following equation:

\begin{equation}
V_{Imp}=\frac{R_{NCA}}{R_{OM}}, \quad V_{Imp}\in [0,1]
\end{equation}

Where $R_{NCA}$ is the ratio of non-compliant attributes for each configuration and $R_{OM}$ is the ratio of the order of magnitude.  $R_{NCA}$ can be obtained by dividing the number of non-compliant attributes ($N_{NCA}$) for each configuration over the total number of attributes ($N_{TA}$) for that configuration. One possible example of calculating this measure is through the equation that can be defined as follows:

\begin{equation}
R_{NCA}=\frac{N_{NCA}}{N_{TA}}, \quad R_{NCA}\in [0,1] 
\end{equation}

On the other hand, $R_{OM}$ represents the order-of-the-magnitude of the impact and it depends on the considered context ($R_{OM}\in [0,1]$). For example, a radio context can be considered. In this case, $R_{OM}$ will be equal to the number of connected UEs at the time of calculation over the maximum number of connected UEs that the network function can withstand. 
Another context example is transport, where the configuration compliance of IPsec tunnels are evaluated. In this case, $R_{OM}$ will be equal to the number of non-compliant IPsec tunnels configurations that are configured on the network function over the configured number of IPsec configurations that can be configured for the network function.
Since $R_{OM}$ and $R_{NCA}$ have decimal values ranging between 0 and 1, their product which represents the vulnerability impact measure will also range between 0 and 1 however medium impact will be shifted towards 0 (majority of the values resulting from the multiplication operation will be concentrated close to 0; the range will be stretched and medium impact would be around 0.25). To shift the medium impact value back to 0.5, a scaling operation can be used. 

An example scaling function can be defined as follows:
\begin{itemize}
\item For values $< 0.25$:
\begin{align}
V_{Imp}&=\frac{V_{Imp}-min_{old}}{max_{old}-min_{old}}\times (max_{new}-min_{new})+min_{new}\\ &= \frac{V_{Imp}-0}{0.25-0}\times (0.5-0)+0 \\ &=\frac{V_{Imp}}{0.25}\times 0.5 
\end{align}
As a result, low impact values will range between $0$ and $0.5$.
\item For values $> 0.25$:
\begin{align}
V_{Imp}&=\frac{V_{Imp}-min_{old}}{max_{old}-min_{old}}\times (max_{new}-min_{new})+min_{new}\\ &= \frac{V_{Imp}-0.25}{1-0.25}\times (1-0.5)+0.5 \\ &=\frac{V_{Imp}-0.25}{0.75}\times 0.5+0.5 
\end{align}
As a result, high impact values will range between $0.5$ and $1$.
\end{itemize}

\textbf{Duration of non-compliant measure:} 
This measure ($D_{NC}$) represents the average patching time of non-compliant rules for a network function. It is updated periodically until patching is done (for every rule). When a patch is done, the corresponding value for a specific rule goes back to zero. $D_{NC}$ can be calculated for each network function using the following:

\begin{equation}
\overrightarrow{Rules_{NC}}=[T_{NC1}, T_{NC2}, \dots]
\end{equation}

For every element in $\overrightarrow{Rules_{NC}}$:

\begin{equation}
T_{NCx}=
\begin{cases}
    min(T_{NCx}+\frac{T_{TP}}{T_{SC}}, 1), & \text{if no compliance} \\
    0, & \text{else}
\end{cases}
\end{equation}

Where $T_{NCx}$  is the interval for each rule to be patched (or remain unpatched), $T_{SC}$ is the configured scanning period, and  $T_{TP}$  is the upper limit of time to patch (e.g., 90 days according to industry best practices on disclosing vulnerabilities).
The measure $D_{NC}$ can then be obtained using:
\begin{equation}
D_{NC}=\frac{\sum \overrightarrow{Rules_{NC}}}{N_{NC}}, \quad D_{NC}\in [0,1]
\end{equation}

Where $N_{NC}$ is the number of non-compliant rules. 

\textbf{Asset criticality measure:}
The asset criticality measure ($A_{Cr}$) is used to reflect the criticality of a network function in a domain, and it can be represented by several variables such as data sensitivity, availability, location, dependency, etc. The selected variables can have binary values (e.g., data sensitivity: sensitive $1$ or non-sensitive $0$). 
After selecting the inputs for the $A_{Cr}$ measure, the total number of possible of combinations can be calculated and the resulting values can be grouped. For example, if three inputs are selected then the total number of possible combinations is $8$ ($2^3$) and the number of groups is $4$ (i.e., group 1: all 0's, group 2: one 1, group 3: two 1's, group 4: all 1's). In this case, each group will have a score. Considering the same example, the four groups will have scores $0$, $0.33$, $0.67$ and $1$, respectively. Note that $0$ is the desired value representing one end of the spectrum (lowest criticality) and $1$ representing the other (highest criticality).     

The above calculation method example (based on three inputs: data sensitivity, availability and location) is presented in Table~\ref{acr}.

\begin{table*}[ht!]
\centering
\begin{tabular}{|c|c|c|c|c|}
\hline
 & Data sensitivity & Availability & Location & Score \\
\hline
Group 1 & 0 (not sensitive)&0 (tolerates delay)&0 (internally exposed) & 0\\\hline
Group 2 & 1 (sensitive)&0 (tolerates delay)&0 (internally exposed) & 0.33 \\
Group 2 & 0 (not sensitive)&1 (high availability)&0 (internally exposed) & 0.33\\
Group 2 &  0 (not sensitive)&0 (tolerates delay)&1 (externally exposed) & 0.33 \\\hline
Group 3 & 0 (not sensitive)&1 (high availability)&1 (externally exposed) & 0.67 \\
Group 3 & 1 (sensitive)&0 (tolerate delay)&1 (externally exposed) & 0.67 \\
Group 3& 1 (sensitive)&1 (high availability)&0 (internally exposed) & 0.67\\\hline
Group 4 & 1 (sensitive)&1 (high availability)&1 (externally exposed) & 0.67\\
\hline
\end{tabular}
\caption{A calculation example of the asset criticality measure}\label{acr}
\end{table*}

\textbf{Environment impact measure:}
This measure ($Env_{Imp}$) can be a function (e.g., average) of the local security states of the neighboring network functions (i.e., directly connected network functions to a specific network function) in a domain.  One possible example of calculating this measure is through the equation that can be defined as follows:

\begin{equation}
Env_{Imp}=\frac{\sum_{1}^{X-1}VulMet_{L,x}^{Human}}{X-1}, \quad Env_{Imp}\in [0,1]
\end{equation}

Where $VulMet_{L,x}^{Human}$ is the local security state of network function $x$, and $X$ is the total number of directly connected network functions. 

\subsection{Attack Surface Exposure Metric}

Attack surface represents a subset of the system's resources that an attacker can utilize to attack the system. The authors in \cite{manadhata2004measuring} defined an attack surface in terms of the system's resources which is composed of the set of entry points the attacker used to conduct their attack (e.g., radio interface), the system channels an attacker used to connect to the system (e.g, control plane protocols), and the data items that were being sent to the system to compromise or threaten the system (e.g. Buffer status reports). In this work we utilized the same definition and formalism of attack surface to introduce a quantified attack surface exposure metric. A formal definition of attack surface can be found below:

\begin{equation}
    AS := \langle E, C, D\rangle
\end{equation}

Where $AS$ represents the attack surface, $E$ refers to the entry point, $C$ represents the channel and $D$ represents the data.

Consequently, having a more granular view on the different entry points of a system will provide eventually a better interpretation of a score that is calculated for the attack surface to understand exactly where is the attack surface that requires prioritization, hence we categorized the attack surfaces present in a 6G NF to be as listed below:

\begin{itemize}
\item \textbf{3GPP Radio Attack Surface:} describes the attack surface within the radio interface
\item \textbf{3GPP Network Attack Surface:} describes the attack surface of the IP-based interfaces defined in 3GPP for example, F1, E1, N2, N3, SBA interfaces etc.
\item \textbf{O-RAN Network Attack Surface:} describes the attack surface of the O-RAN interfaces in a O-RAN architecture, for example E2, A1, O1, O2.
\item \textbf{Operations and Management (OAM) Attack Surface:} describes the attack surface of the interfaces that are used for the OAM purpose.
\item \textbf{Platform Attack Surface:} describes the attack surface on the platform layer that can include the underlying host (i.e OS), the orchestration layer (i.e K8s), or other supporting services.
\end{itemize}

The attack surface exposure metric is a measure of how the attack surface (e.g., entry points, channel, data) of network function has increased or decreased due to changes in the network or the network function's environment, triggered by events. This metric can be used individually to understand the causal relationships and the security impact of changes in the network. Taking RAN as an example with a focus on the 3GPP radio attack surface, a cell represents the entry point, and the channel represents the radio protocols while data are the functions, procedures, and methods that can be invoked within each protocol and leveraged to attack a radio system. Generally, the attack surface exposure metric can be calculated through the equation defined below:

\begin{equation}
    AS_{E} = (\sum_{i=0}^{e}D_{i} \times OM_{EPi}) \times \frac{EP_{C}}{EP_{Max}}, \quad AS_{E}\in [0,1]
\end{equation}

$AS_{E}$ represents the attack surface exposure metric, and $\sum_{i=0}^{e}D_{i}$ is the sum of possibly attacked data items in an attack surface per entry point $e$. $OM_{EPi}$ represents the order of magnitude in one entry point. The order of magnitude is a context-based value that is related to the resource that utilizes the data components of the attack surface. Referring to the radio interface as an example, the order of magnitude can be the ratio of potential attacking UEs to the total number of supported UE per cell (entry point). The order of magnitude can be represented with the below equation:

\begin{equation}
    OM_{R} = \frac{UE_{PA}}{UE_{Connected}}
\end{equation}

Where $UE_{PA}$ represents the current number of potential attacking UEs, and $UE_{Connected}$  is the total number of the current connected UEs in a cell at the time of the metric calculation. Moreover, $EP_{C}$ represents the currently configured number of entry points (e.g., the number of configured cells) and $EP_{Max}$ is the maximum number of allowed entry points (e.g., the maximum number of cells as defined as part of the capacity of the DU, CU, or gNB).

\section{Related Work} \label{sec:related}
There has been efforts done in the area of measuring security in both academia and industry. In this paper, we attempted to analyze both aspects to highlight the advancements that we provide to complement existing work. 

Industrial solutions commonly focus on a single metric in order to provide a quantifiable score for the security posture of a system or a service. The existing solutions focus on measuring vulnerabilities by analyzing their criticality and exploitability with  software vulnerabilities being the focal point. Another method that is used, is measuring the compliance of configuration with a secure configuration. The existing solutions did not address combining different metrics to understand the posture of the system. An example of leveraging configuration compliance as a method of measuring security is used by \cite{mcs}, where the solution relates the security posture to a benchmark checklist to be followed. This approach provides an incomplete view of the security posture, since it is static and does not evolve with changes of the environment and its requirements. Other solutions like \cite{autobahn} focuses on relating the security state evaluation to the exploitability of vulnerabilities, which can be useful but in order to understand the corrective action to be taken, there is a need for an additional understanding on what type of controls that exist and their performance. 

Academic literature for measuring security in complex systems such as mobile networks present several problems. Perhaps the most pressing issue is the subjectivity of the risk assessment process, as highlighted in \cite{martin2008making} and \cite{noll2014measurable}. Reliance on expert judgment to evaluate risks can lead to inconsistencies and biases in risk assessment, making it difficult to develop effective and actionable steps to address security risks. Another issue is the lack of specificity in identifying metrics and measures for quantifying and comparing security risks, as observed in \cite{sahinoglu2008input}. The absence of well-defined security scores can make it challenging to develop a comprehensive and adaptable approach to manage the risks. Additionally, excluding impact of the network as in \cite{sahinoglu2008input}, an essential factor in distributed systems where the actions of different components can have far-reaching implications beyond the node itself.

The work done in \cite{shafayat2022assessing} has taken 5G networks as its domain, and it focused on defining vulnerability metrics and attack surface metrics to evaluate a security posture of a 5G network function. However, it did not consider the impact of existence of security controls on the security state. The security controls that are implemented for mitigation are a crucial factor for evaluating a security posture.

Perhaps the survey done in \cite{pendleton2016survey} is the closest to this paper. The work has identified the different metrics that can be an input to calculate and evaluate a security state. However, some of the metrics are subjective and do not serve the purpose of automation. For example, there is an identified metric that measures attacks, where attacks can be zero-day, targeted attacks, malware spreading. The drawback of this type of measures that it relies on a human analyst that inspects different data sources to derive a numerical conclusion. The process is prone to error and cannot be automated, thus there is a risk to provide a false sense of security.

To our knowledge, the studied literature and state of the art do not take into consideration the factors that induce changes to the security state. Additionally, existing work do not consider the effect of the deployment of security controls and how their effectiveness can impact the security state. Moreover,the reliance on subjectivity when calculating metrics or assigning a score. Metrics needs to be objective, subjectivity is a hindering characteristic of the metric and is an obstacle to reuse the metric in different deployments with different calculating methods.

Our Contribution is that we provide an approach based on a defined security state model that is used to identify the metrics that affect the security state and the metrics relationship. Additionally, the provided metrics and their sub-metrics were defined with properties that ensures its usefulness, for example the defined metrics are objective, dynamic, comparable, can be automated and of course they are quantitative with scalar values that have no units.

\section{Security Metrics Enabling Intent-based Security Management} \label{sec:usage}

Establishing security metrics in 5G and 6G deployments, as presented in Section~\ref{sec:metrics}, lays the foundation for zero-touch security automation capabilities that hinges on the observability of such metrics. The availability of ready to use quantifiable metrics enable more efficient decision-making with less human intervention. Additionally, it provides information about situational awareness of the environment and the security posture of 6G nodes in the network. %which in turn facilitates risk-based decisions, as the calculated metrics offers a simple way to rank security issues per their impact.

\begin{figure}[t!]
    \includegraphics[width=0.98\linewidth]{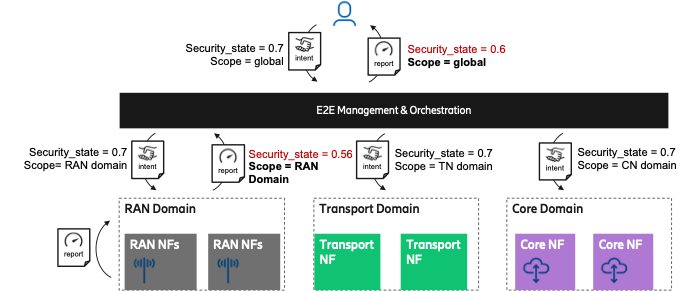}
    \caption{Security metrics supporting intent-based security management}
    \label{fig:metric_ibm}
\end{figure}

Security metrics are essential enablers for an intent-based security management. In its core essence, intent-based management networks as defined by TMForum and ETSI ZSM \cite{gomes2021intent} is an autonomous element, where it adopts closed loop automation. The MAPE-K, as a closed loop automation architecture \cite{computing2006architectural}, has its first phase that as the Monitoring phase, which is responsible for collecting, processing and observing metrics and KPIs that are provided to the analysis phase afterwards. Thus, having the right metrics and KPIs is an essential component to achieve the automation level that is promised by autonomous elements which is the driving component of a zero-touch network and security in 6G networks \cite{ericsson6gautomation}. 

As illustrated in Figure \ref{fig:metric_ibm}, it shows a high level view of an intent-based security management loop. It is shown that on the highest level a human operator or the intent owner is responsible on defining the security requirements, in this case the security requirement that is set is to maintain the Security State at a specific quantifiable score, 70\% and that this security state will be applied globally on all domains, RAN, Transport and Core. The security requirement and scope is pushed towards the E2E management and orchestration which is the layer responsible on decomposing the high level business intent into the lower level intents, each new intent will be then pushed towards its respective network domain. 

Each domain has its intent handler that is responsible on decomposing further the intents into either further intent or low level actuation that will be acted on the managed resource. The intent handler will be responsible on collecting data and monitoring the metrics that were discussed in the previous sections, consequently it will evaluate if the output of the metrics calculations met the requirements in the intent or not, if they meet then it will report back the score and that the intent requirement has met, as shown in the transport and core domains. However, if the requirement is not met like what is illustrated in the RAN domain, the intent handler will report to the upper layers the reason the score is low and which metric has most impact on violating the intent. %For example, the intent report can include that the most impacting metric is the attack exposure metric, thus it provides an explanation to the highest layer where the effort should be focused on.%The intent handler has an understanding of the translation between a requirement of a security state in its domain will be able then to translate those requirements into actual low level executions. Where each intent handler implements its own closed security automated loop. 

\section{Conclusions and Future Work} \label{sec:conclusion}
This paper addressed the problem of measuring the security state of mobile telecommunication systems. As future networks (i.e 6G) evolves towards automated and intent-based management, due to the challenges of increased complexity, heterogeneity and scale of deployments, measurability of security becomes a cornerstone. Being able to quantify and measure security objectively enables better decision-making, informed risk management, and enhances the trust in deployed networks that are less error-prone via zero-touch configuration.

%In support of such requirements, we have developed a state machine model to capture the security life cycle of deployed network functions and the transitions between the different states comprising the life cycle. By leveraging this model, security metrics can support the evaluation of transitions and enable the calculation of an overall security score.

%We have identified three essential security metrics that make up the security state score: 1) the \textit{exposure of the attack surface}, 2) the \textit{impact of system vulnerabilities}, and 3) the \textit{effectiveness and coverage of applied security controls}. This comprehensive view of metrics that impact a security state, provides a better understanding of the security posture of a system than just looking at a single type of metrics in isolation (e.g., vulnerabilities, or exposure). 
%We advocate that this work lays the foundation for achieving effective security management in 6G.

%Lastly, we have opened up on how to apply practically this model to a 6G deployment, illustrating how it can support the development of automation intent-based security management and zero-touch security.

In future work, we will conduct empirical experiments with various use cases to validate the mathematical models. Moreover, we will attempt to use the metrics for designing and developing a closed security automation loop where the loop will be managed by intents and the metrics will be the requirements defined in an intent model.

Finally, there are yet challenges to be explored in this area. One challenge related to the mathematical nature of the output score which is uncertainty and the error allowance. Additionally, another challenge is to be able to explain the score to provide insights for humans supervising the automation loop.
%Another exploration is to identify the security requirements (i.e., SLAs) that can be set for network slice security and what are the suitable metrics from those defined in this work can be used to evaluate a security state of a network slice to ensure the fulfillment of the SecSLA that was set in the slice profile.

%Finally, we will be investigating conflict management between the security requirements that are observed by the metrics and other network or computational constraints.

\bibliographystyle{bib/IEEEtran}
\bibliography{bbl/conference_101719}

@TechReport{ 3gpp33501, 
    author = "3GPP", 
    title = "5G;Security architecture and procedures for 5G System", 
    year = "2024", 
    number = "TS 33.501 version 18.5.0 Release 18" 
}


@online{gsmathreats,
    author = "GSMA",
    title = "Mobile Telecommunications Threat Landscape 2023",
    url  = "https://www.gsma.com/security/wp-content/uploads/2023/02/GSMA-Mobile-Telecommunications-Security-Landscape-2023\_v1\_for-website.pdf",
    addendum = "(accessed: 03.05.2024)"
}

@online{autobahn,
    author = "AutoBahn",
    url  = "https://autobahn-security.com",
    addendum = "(accessed: 03.05.2024)"
}

@online{mcs,
    author = "Microsoft",
    title = "Microsoft Secure Score",
    url  = "https://learn.microsoft.com/en-us/azure/defender-for-cloud/secure-score-security-controls",
    addendum = "(accessed: 03.05.2024)"
}


@online{ericsson6gautomation,
    author = {Ericsson},
    title = {How Zero Can We Touch},
    url  = "https://www.ericsson.com/en/blog/2023/4/how-zero-can-we-touch",
    addendum = "(accessed: 03.05.2024)"
}

@online{ericsson6gspectrum,
    author = {Ericsson},
    title = {6G spectrum - enabling the future mobile life beyond 2030},
    url  = "https://www.ericsson.com/en/reports-and-papers/white-papers/6g-spectrum-enabling-the-future-mobile-life-beyond-2030",
    addendum = "(accessed: 03.05.2024)"
}

.

@inproceedings{khan2018aes,
  title={{AES} and {SNOW 3G} are Feasible Choices for a {5G} Phone from Energy Perspective},
  author={Khan, Mohsin and Niemi, Valtteri},
  booktitle={5G for Future Wireless Networks: First International Conference, 5GWN 2017, Beijing, China, April 21-23, 2017, Proceedings 1},
  pages={403--412},
  year={2018},
  organization={Springer}
}

@inproceedings{ekdahl2021snow,
  title={{SNOW-Vi}: an extreme performance variant of {SNOW-V} for lower grade CPUs},
  author={Ekdahl, Patrik and Maximov, Alexander and Johansson, Thomas and Yang, Jing},
  booktitle={Proceedings of the 14th ACM Conference on Security and Privacy in Wireless and Mobile Networks},
  pages={261--272},
  year={2021}
}

@article{computing2006architectural,
  title={An architectural blueprint for autonomic computing},
  author={Computing, Autonomic and others},
  journal={IBM White Paper},
  volume={31},
  number={2006},
  pages={1--6},
  year={2006},
  publisher={Citeseer}
}

@article{gomes2021intent,
  title={Intent-driven closed loops for autonomous networks},
  author={Gomes, Pedro Henrique and Buhrgard, Magnus and Harmatos, J{\'a}nos and Mohalik, Swarup Kumar and Roeland, Dinand and Niem{\"o}ller, J{\"o}rg},
  journal={Journal of ICT Standardization},
  volume={9},
  number={2},
  pages={257--290},
  year={2021},
  publisher={River Publishers}
}
@inproceedings{harrilal2023performance,
  title={Performance Evaluation of Quantum-Resistant Open Fronthaul Communications in 5G},
  author={Harrilal-Parchment, Ricardo and Pujol, Isabela Fernandez and Akkaya, Kemal},
  booktitle={IEEE INFOCOM 2023-IEEE Conference on Computer Communications Workshops (INFOCOM WKSHPS)},
  pages={1--6},
  year={2023},
  organization={IEEE}
}

@article{groen2024securing,
  title={Securing O-RAN Open Interfaces},
  author={Groen, Joshua and D'Oro, Salvatore and Demir, Utku and Bonati, Leonardo and Villa, Davide and Polese, Michele and Melodia, Tommaso and Chowdhury, Kaushik},
  journal={IEEE Transactions on Mobile Computing},
  year={2024},
  publisher={IEEE}
}

@inproceedings{martin2008making,
  title={Making security measurable and manageable},
  author={Martin, Robert A},
  booktitle={MILCOM 2008-2008 IEEE Military Communications Conference},
  pages={1--9},
  year={2008},
  organization={IEEE}
}

@article{noll2014measurable,
  title={Measurable security, privacy and dependability in smart grids.},
  author={Noll, Josef and Garitano, Inaki and Fayyad, Seraj and {\AA}sberg, Erik and Abie, Habtamu},
  journal={J. Cyber Secur. Mobil.},
  volume={3},
  number={4},
  pages={371--398},
  year={2014}
}

@article{sahinoglu2008input,
  title={An input--output measurable design for the security meter model to quantify and manage software security risk},
  author={Sahinoglu, Mehmet},
  journal={IEEE Transactions on Instrumentation and Measurement},
  volume={57},
  number={6},
  pages={1251--1260},
  year={2008},
  publisher={IEEE}
}

@phdthesis{shafayat2022assessing,
  title={Assessing Security in the Multi-Stakeholder Premise of 5G: A Survey and an Adapted Security Metrics Approach},
  author={Shafayat Oshman, Muhammad},
  year={2022},
  school={Carleton University}
}

@article{pendleton2016survey,
  title={A survey on systems security metrics},
  author={Pendleton, Marcus and Garcia-Lebron, Richard and Cho, Jin-Hee and Xu, Shouhuai},
  journal={ACM Computing Surveys (CSUR)},
  volume={49},
  number={4},
  pages={1--35},
  year={2016},
  publisher={ACM New York, NY, USA}
}

@article{coronado2022zero,
  title={Zero touch management: A survey of network automation solutions for 5G and 6G networks},
  author={Coronado, Estefan{\'\i}a and Behravesh, Rasoul and Subramanya, Tejas and Fern{\`a}ndez-Fern{\`a}ndez, Adriana and Siddiqui, Muhammad Shuaib and Costa-P{\'e}rez, Xavier and Riggio, Roberto},
  journal={IEEE Communications Surveys \& Tutorials},
  volume={24},
  number={4},
  pages={2535--2578},
  year={2022},
  publisher={IEEE}
}

@techreport{schroeder2022performance,
  title={Performance Measurement Guide for Information Security},
  author={Schroeder, Katherine and Trinh, Hung},
  year={2022},
  institution={National Institute of Standards and Technology}
}

@book{manadhata2004measuring,
  title={Measuring a system's attack surface},
  author={Manadhata, Pratyusa and Wing, Jeannette Marie},
  year={2004},
  publisher={School of Computer Science, Carnegie Mellon University Pittsburgh, PA, USA}
}



\begin{thebibliography}{10}
\providecommand{\url}[1]{#1}
\csname url@samestyle\endcsname
\providecommand{\newblock}{\relax}
\providecommand{\bibinfo}[2]{#2}
\providecommand{\BIBentrySTDinterwordspacing}{\spaceskip=0pt\relax}
\providecommand{\BIBentryALTinterwordstretchfactor}{4}
\providecommand{\BIBentryALTinterwordspacing}{\spaceskip=\fontdimen2\font plus
\BIBentryALTinterwordstretchfactor\fontdimen3\font minus \fontdimen4\font\relax}
\providecommand{\BIBforeignlanguage}[2]{{%
\expandafter\ifx\csname l@#1\endcsname\relax
\typeout{** WARNING: IEEEtran.bst: No hyphenation pattern has been}%
\typeout{** loaded for the language `#1'. Using the pattern for}%
\typeout{** the default language instead.}%
\else
\language=\csname l@#1\endcsname
\fi
#2}}
\providecommand{\BIBdecl}{\relax}
\BIBdecl

\bibitem{coronado2022zero}
E.~Coronado, R.~Behravesh, T.~Subramanya, A.~Fern{\`a}ndez-Fern{\`a}ndez, M.~S. Siddiqui, X.~Costa-P{\'e}rez, and R.~Riggio, ``Zero touch management: A survey of network automation solutions for 5g and 6g networks,'' \emph{IEEE Communications Surveys \& Tutorials}, vol.~24, no.~4, pp. 2535--2578, 2022.

\bibitem{ericsson6gspectrum}
\BIBentryALTinterwordspacing
Ericsson. 6g spectrum - enabling the future mobile life beyond 2030. [Online]. Available: \url{https://www.ericsson.com/en/reports-and-papers/white-papers/6g-spectrum-enabling-the-future-mobile-life-beyond-2030}
\BIBentrySTDinterwordspacing

\bibitem{schroeder2022performance}
K.~Schroeder and H.~Trinh, ``Performance measurement guide for information security,'' National Institute of Standards and Technology, Tech. Rep., 2022.

\bibitem{3gpp33501}
3GPP, ``5g;security architecture and procedures for 5g system,'' Tech. Rep. TS 33.501 version 18.5.0 Release 18, 2024.

\bibitem{khan2018aes}
M.~Khan and V.~Niemi, ``{AES} and {SNOW 3G} are feasible choices for a {5G} phone from energy perspective,'' in \emph{5G for Future Wireless Networks: First International Conference, 5GWN 2017, Beijing, China, April 21-23, 2017, Proceedings 1}.\hskip 1em plus 0.5em minus 0.4em\relax Springer, 2018, pp. 403--412.

\bibitem{ekdahl2021snow}
P.~Ekdahl, A.~Maximov, T.~Johansson, and J.~Yang, ``{SNOW-Vi}: an extreme performance variant of {SNOW-V} for lower grade cpus,'' in \emph{Proceedings of the 14th ACM Conference on Security and Privacy in Wireless and Mobile Networks}, 2021, pp. 261--272.

\bibitem{gsmathreats}
\BIBentryALTinterwordspacing
GSMA. Mobile telecommunications threat landscape 2023. [Online]. Available: \url{https://www.gsma.com/security/wp-content/uploads/2023/02/GSMA-Mobile-Telecommunications-Security-Landscape-2023\_v1\_for-website.pdf}
\BIBentrySTDinterwordspacing

\bibitem{manadhata2004measuring}
P.~Manadhata and J.~M. Wing, \emph{Measuring a system's attack surface}.\hskip 1em plus 0.5em minus 0.4em\relax School of Computer Science, Carnegie Mellon University Pittsburgh, PA, USA, 2004.

\bibitem{mcs}
\BIBentryALTinterwordspacing
Microsoft. Microsoft secure score. [Online]. Available: \url{https://learn.microsoft.com/en-us/azure/defender-for-cloud/secure-score-security-controls}
\BIBentrySTDinterwordspacing

\bibitem{autobahn}
\BIBentryALTinterwordspacing
AutoBahn. [Online]. Available: \url{https://autobahn-security.com}
\BIBentrySTDinterwordspacing

\bibitem{martin2008making}
R.~A. Martin, ``Making security measurable and manageable,'' in \emph{MILCOM 2008-2008 IEEE Military Communications Conference}.\hskip 1em plus 0.5em minus 0.4em\relax IEEE, 2008, pp. 1--9.

\bibitem{noll2014measurable}
J.~Noll, I.~Garitano, S.~Fayyad, E.~{\AA}sberg, and H.~Abie, ``Measurable security, privacy and dependability in smart grids.'' \emph{J. Cyber Secur. Mobil.}, vol.~3, no.~4, pp. 371--398, 2014.

\bibitem{sahinoglu2008input}
M.~Sahinoglu, ``An input--output measurable design for the security meter model to quantify and manage software security risk,'' \emph{IEEE Transactions on Instrumentation and Measurement}, vol.~57, no.~6, pp. 1251--1260, 2008.

\bibitem{shafayat2022assessing}
M.~Shafayat~Oshman, ``Assessing security in the multi-stakeholder premise of 5g: A survey and an adapted security metrics approach,'' Ph.D. dissertation, Carleton University, 2022.

\bibitem{pendleton2016survey}
M.~Pendleton, R.~Garcia-Lebron, J.-H. Cho, and S.~Xu, ``A survey on systems security metrics,'' \emph{ACM Computing Surveys (CSUR)}, vol.~49, no.~4, pp. 1--35, 2016.

\bibitem{gomes2021intent}
P.~H. Gomes, M.~Buhrgard, J.~Harmatos, S.~K. Mohalik, D.~Roeland, and J.~Niem{\"o}ller, ``Intent-driven closed loops for autonomous networks,'' \emph{Journal of ICT Standardization}, vol.~9, no.~2, pp. 257--290, 2021.

\bibitem{computing2006architectural}
A.~Computing \emph{et~al.}, ``An architectural blueprint for autonomic computing,'' \emph{IBM White Paper}, vol.~31, no. 2006, pp. 1--6, 2006.

\bibitem{ericsson6gautomation}
\BIBentryALTinterwordspacing
Ericsson. How zero can we touch. [Online]. Available: \url{https://www.ericsson.com/en/blog/2023/4/how-zero-can-we-touch}
\BIBentrySTDinterwordspacing

\end{thebibliography}

\end{document}